\documentclass[prd,aps,preprint,tightenlines,superscriptaddress]{revtex4}
\usepackage{epsfig}
\usepackage{amsmath}
\usepackage{slashed}

\begin{document}


\title{QCD Factorization for Deep-Inelastic Scattering \\
At Large Bjorken $x_B \sim 1-{\cal O}( \Lambda_{\rm QCD}/Q)$}

\author{Panying Chen}
\affiliation{Department of Physics, University of Maryland, College
Park, Maryland 20742, USA}
\author{Ahmad Idilbi}
\affiliation{Department of Physics, University of Maryland,
College Park, Maryland 20742, USA}
\author{Xiangdong Ji}
\affiliation{Department of Physics, University of Maryland, College
Park, Maryland 20742, USA} \affiliation{Department of Physics,
Peking University, Beijing, 100871, P. R. China}

\date{\today}
\vspace{0.5in}
\begin{abstract}
We study deep-inelastic scattering factorization on a nucleon in
the end-point regime $x_B \sim 1-{\cal O}( \Lambda_{\rm QCD}/Q)$
where the traditional operator product expansion is supposed to fail.
We argue, nevertheless, that the standard result holds to
leading order in $1-x_B$ due to the absence of the scale
dependence on $(1-x_B)Q$. Refactorization of the scale $(1-x_B)Q^2$
in the coefficient function can be made in the
soft-collinear effective theory and remains valid in the end-point
regime. On the other hand, the traditional refactorization approach
introduces the spurious scale $(1-x_B)Q$ in various factors, which
drives them nonperturbative in the region of our interest.
We show how to improve the situation by introducing a
rapidity cut-off scheme, and how to recover the effective
theory refactorization by choosing appropriately the cut-off parameter.
Through a one-loop calculation, we demonstrate explicitly that
the proper soft subtractions must be made in the collinear
matrix elements to avoid double counting.

\end{abstract}
\maketitle

\section{Introduction}

In lepton-nucleon deep-inelastic scattering (DIS), the Bjorken
regime with virtual photon mass $Q^2\rightarrow \infty$ and
$x_B=Q^2/2M\nu$ fixed presents a textbook example of
perturbative QCD (pQCD) factorization \cite{collins89}. In this
regime, the scale $(1-x)^\alpha Q$ goes to infinity (we drop the
subscript $B$ on $x$ henceforth), where $\alpha$ is any real number,
or at least much larger than the soft QCD scale $\Lambda_{\rm
QCD}$. An alternative DIS regime is $Q^2\rightarrow \infty$ with
$(1-x)Q \sim \Lambda_{\rm QCD}$, where the final hadron state
invariant mass $(1-x)Q^2\sim Q\Lambda_{\rm QCD}$ is still large
and is distinct from the resonance region. This large-$x$
regime has so far received little attention in theory, possibly
because it covers only a small kinematic interval in real
experiments. The existing QCD studies in the literature are
somewhat controversial \cite{kim,pecjak}.

In this paper, we present a factorization study of DIS at this
regime. The main point we advocate here is that the standard
pQCD factorization remains valid in this new kinematic domain
to leading order in $1-x$.
Then we move on to discuss refactorization which factorizes the
physics at scale $(1-x)Q^2$ from that at $Q^2$. The useful
theoretical framework for this purpose is the so-called
soft-collinear effective field theory (SCET) developed recently
\cite{SCET}. Indeed, the first treatment of the large-$x$ region
in DIS using SCET was made in \cite{manohar} and followed
in \cite{kim,pecjak} for $(1-x)Q\sim \Lambda_{\rm QCD}$
[see also \cite{neu}.] Because the scale $(1-x)Q$ does not enter in
the perturbative calculation, the final result amounts to a
standard pQCD factorization, with the additional benefit that the
refactorization becomes manifest. One subtlety we discuss
extensively in this paper is the role of the soft contribution and
its relation to the light-cone parton distribution. In a recent paper
\cite{kim}, a factorization formula for the DIS structure function
is derived in SCET, which is similar to what we find here.
However, because of the lack of a consistent regularization and
clear separation of contributions among different factors, the result
does not recover that of Ref.~\cite{manohar} in the limit $(1-x)Q\gg
\Lambda_{QCD}$.

The traditional approach of refactorization was pioneered in \cite{sterman},
where a new parton distribution together
with soft factor and jet function is introduced. These matrix elements
are designed to absorb large logarithms generated from soft gluon
radiations off on-shell lines. Evolution equations for them are derived and
solved to resum the large soft-gluon logarithms. In this approach, the
factorization of scales are not apparent from the start.
Moreover, a spurious scale $(1-x)Q$ appears in all factors which
makes them nonperturbative in the regime of our
interest. After reviewing this, we present a more general
factorization along this line with dependence on a rapidity cut-off
$\rho$. Different choices of the cutoff lead to
redistributions of large logarithms in different matrix elements.
A particular choice yields a picture similar to that of the SCET
approach.

The presentation of this paper is as follows. In section II, we
argue that the standard factorization approach remains valid in the end-point
regime $x \sim 1-{\cal O}( \Lambda_{\rm QCD}/Q)$. In section III, we present
an effective field theory approach to refactorization, following the previous
work of Ref. \cite{kim}. The difference is that our result is consistent
with that of Ref. \cite{manohar}, with the jet
factor absorbing all physics at the intermediate scale $(1-x)Q^2$.
We explain that the soft and collinear contributions
combine to give the light-cone parton distribution. In section IV,
we first review the traditional factorization in which various
matrix elements are introduced to account for soft gluon radiations. We
then show how to derive a more general factorization
with a rapidity cutoff. We demonstrate explicitly that
the proper soft subtractions must be made in the collinear
matrix elements to avoid double counting.
By choosing different cut-off, we find different
pictures of factorization and large-logarithmic resummation.
The effective field theory refactorization can be recovered this way.

\section{Validity of the standard QCD factorization at $x_B \sim 1-{\cal O}(
\Lambda_{\rm QCD}/Q)$ }

The standard pQCD factorization theorem is derived in the Bjorken
limit in which $Q^2\rightarrow \infty$ and $x$ is a fixed constant
between $0$ and $1$. To leading order in $1/Q^2$, the proton's
spin-independent structure function $F_1(x, Q^2)$ can be factorized
as
\begin{equation}
   F_1(x, Q^2) = \sum_f \int^1_x \frac{dy}{y}
       C_f\left(\frac{x}{y}, \frac{Q^2}{\mu^2} \right) q_f(y, \mu^2) \ ,
\end{equation}
where $\mu$ is a factorization scale, $C_f$ is the coefficient
function depending on scale $Q^2$ and $\mu^2$ (factorization scale),
and $q_f$ is a quark distribution of flavor $f$. For simplicity, we
omit the quark charges and gluon contribution which are inessential
for our discussion. In the moment space, the factorization takes a
simple product form,
\begin{equation}
   F_1^N(Q^2) = \sum_f C_{fN}(Q^2/\mu^2) q_{fN}(\mu^2) \ ,
\end{equation}
where the moments are defined the usual way, e.g., $q_N = \int^1_0
dxx^{N-1}(q(x)+\bar q(x))$ with $N$ even.

The above factorization in principle is invalid in non-Bjorken
regions. However, we argue that it still holds in the regime of
our interest, $(1-x)Q \sim \Lambda_{\rm QCD}$, with the same
physical parton distributions to leading order in $1-x$. The main
point is that although we now have a new infrared scale $(1-x)Q$,
it does not appear in the above factorization.

Indeed, as $x\rightarrow 1$ the coefficient function $C_f$ can
only depend on the two hard scales---invariant photon mass $Q^2$
and the final-hadron-state invariant mass $(1-x)Q^2$, both remain
large when $(1-x)Q\sim \Lambda_{\rm QCD}$. Hence there is no
emerging infrared scale entering physical observables in this new
regime, and the original factorization remains valid to
leading order in $1-x$.

The above observation can be seen clearly in the one-loop result.
The coefficient function at order $\alpha_s$ in the
$x$-space
\begin{equation}
   C^{(1)}(x)=\frac{\alpha_s}{2\pi}C_F\left[
\left(\frac{2\ln((1-x)Q^2/\mu^2)
    - 3/2}{1-x}\right)_+ - \left(\frac{3}{2} \ln\frac{Q^2}{\mu^2} +
\frac{9}{2}
      + \frac{\pi^2}{3} \right)      \delta(1-x)\right] \ ,
\end{equation}
where we have neglected higher order in $(1-x)$. The scheme we use
here is the modified minimal subtraction (${\overline {\rm MS}}$).
In term of moments, one finds
\begin{equation}
   C_N^{(1)}\left(\frac{Q^2}{\mu^2},\frac{Q^2}{N\mu^2}\right) = \frac{\alpha_s}{2\pi}
      C_F\left[\ln^2\frac{Q^2}{N\mu^2}
       - \frac{3}{2}\ln \frac{Q^2}{N\mu^2}- \ln^2\frac{Q^2}{\mu^2}
        + 3 \ln\frac{Q^2}{\mu^2} -
       \frac{9}{2}
        - \frac{\pi^2}{6}\right] \ .
        \label{ff}
\end{equation}
The scale dependence is manifest: The first two logarithms come from
physics at scale $\mu^2 = Q^2/N$, whereas the next two logarithms
come from scale $Q^2$. Clearly, there is no physics from scale
$\mu^2\sim (1-x)^2Q^2\sim Q^2/N^2$. Therefore, even when $(1-x)Q$
becomes of order $\Lambda_{\rm QCD}$, the coefficient function has
no infrared sensitivity to it.

The fundamental reason for the absence of the scale $(1-x)Q$ in a
physical observable is that it is not a Lorentz scalar, whereas
$(1-x)Q^2$ is the invariant mass of the final hadron state. In
principle, the energy of soft gluons and quarks in the Breit frame is
of order $(1-x)Q$ which can appear in the factorization. However,
this happens only for frame-dependent factorization. The
factorization we quoted above is frame-independent and thus any
non-Lorentz scalar cannot appear.

Although the above conclusion appears simple and natural, we have
not seen it stated explicitly in the literature.

\section{Refactorization: Effective Theory Approach}

In the large-$x$ region, independent of whether $(1-x)Q \gg
\Lambda_{\rm QCD}$ or $\sim \Lambda_{\rm QCD}$, there is an
emerging ``infrared" scale $(1-x)Q^2\ll Q^2$. Of course, we always
assume $(1-x)Q^2\gg \Lambda_{\rm QCD}^2$. The presence of this new
scale suggests a further factorization in which the physics
associated with scales $Q^2$ and $(1-x)Q^2$ is disentangled. This
type of factorization was proposed by G. Sterman and others for
the purpose of summing over the large double logarithms of type
$\alpha_s^k [\ln^{i}(1-x)/(1-x)]_+$ $(i\le 2k-1)$ in the
coefficient functions \cite{sterman}. We consider this
refactorization in this and the following sections, commenting on
its applicability in the region of our interest.

We first study refactorization in the effective field theory
(EFT) approach in this section, and will discuss a more intuitive
approach in the next. The EFT method is based on strict scale
separation, very much like the usual QCD factorization discussed
in the previous section. When the scales are separated, one can
sum over large logarithms by using renormalization group
evolutions between scales. Some of the basic discussions here
follow Refs. \cite{manohar,ji, kim}.

\subsection{EFT Refactorization}

To understand the EFT factorization heuristically, we write the
one-loop coefficient function in Eq. (\ref{ff}) in a factorized
form,
\begin{equation}
     C_N^{(1)} \left(\frac{Q^2}{\mu^2},\frac{Q^2}{\overline{N}\mu^2}\right) =
    2  C^{(1)}
\left(\frac{Q^2}{\mu^2}\right) +
     {\cal M}_N^{(1)}\left(\frac{Q^2}{{\overline N}\mu^2}\right)\ ,
\end{equation}
where $C^{(1)}$ is $N$-independent and comes from physics at scale
$Q^2$ and ${\overline N}=Ne^{\gamma_E}$ where $\gamma_E$ is Euler
constant. The one-loop result for $C$ is
\begin{equation}
  C^{(1)} \left(\frac{Q^2}{\mu^2}\right)= \frac{\alpha_s}{4\pi}
C_F\left[-\ln^2\frac{Q^2}{\mu^2}
    + 3 \ln \frac{Q^2}{\mu^2} - 8 + \frac{\pi^2}{6}  \right] \ ,
    \label{C}
\end{equation}
where the constant term is, in principle, arbitrary; we choose it
to be consistent with the effective current below. The two-loop
result for $C$ can be found in \cite{feng,idi}. The other factor
${\cal M}_N^{(1)}$ comes entirely from physics at scale $(1-x)Q^2
\sim Q^2/\overline{N}$,
\begin{equation}
    {\cal M}_N^{(1)}\left(\frac{Q^2}{{\overline N}\mu^2}\right) =
     \frac{\alpha_s}{2\pi}C_F\left[
     \ln^2 \frac{Q^2}{{\overline N}\mu^2}
        - \frac{3}{2} \ln\frac{Q^2}{{\overline N}\mu^2} + \frac{7}{2}
         - \frac{\pi^2}{3}\right] \ ,
\end{equation}
and the second order result for ${\cal M}_N$ can also be found in
\cite{idi}. The key point is that the above refactorization of
scales works to all orders in perturbation theory and EFT provides
a formal approach to establish this: The physics at scale $Q^2$
can be included entirely in $\vert C\vert ^2$ and that at the
other scale is in ${\cal M}_N$.

To arrive at the above refactorization, we start off at the scale
$Q^2$ at which perturbative physics involves virtual gluon
corrections to the hard interaction photon vertex. Note that the
soft-gluon radiations off a hard vertex are usually high-order
effects in $1/Q^2$ and can be neglected. Thus the physics at $Q^2$
can be found from just the quark electromagnetic form factor.
Integrating out physics at scale $Q^2$ is equivalent to matching the
full QCD electromagnetic current to an effective one involving just
soft-collinear physics.
\begin{equation}
     J_{\rm QCD}^{\mu} = C(Q^2/\mu^2) J_{\rm eff}^{\mu}(Q^2/\mu^2)\ ,
\end{equation}
 with the one-loop result given in Eq. (\ref{C}).

We can run the effective current from scale $Q^2$ to scale
$(1-x)Q^2$ using the renormalization group equation
\begin{equation}
      \mu \frac{dJ_{\rm eff}(Q^2/\mu^2)}{d\mu} = -\gamma_1(\alpha_s(\mu)) J_{\rm
      eff}(Q^2/\mu^2) \ ,
\end{equation}
where the anomalous dimension can be calculated from $C$, $\gamma_1
= \mu d\ln C/d \mu$, and has the following generic form,
\begin{equation}
    \gamma_1 = A(\alpha_s) \ln Q^2/\mu^2 + B_1(\alpha_s) \ ,
\end{equation}
in which $A$ and $B_1$ are a series in strong coupling constant $\alpha_s$
and are now known up to three loops \cite{dd}.

At scale $\mu_I^2=(1-x)Q^2$, we follow Ref. \cite{kim}, matching
products of the effective currents to a product of the jet
function, collinear parton contribution, and soft distribution in
SCET. Introducing a small expansion parameter $\lambda$, with
$\lambda^2 Q \sim \Lambda_{\rm QCD}$, $1-x$ in the region of our
interest scales like $\lambda^2$. The collinear partons at the
matching scale $\mu_I^2$ have momentum $(p^+,p^- p_\perp) \sim
Q(1,\lambda^4, \lambda^2)$ [our notation for light-cone components
for arbitrary four-vector $l$ is $l\equiv (l^+,l^-,l_\perp)$ with
$l^{\pm}=\frac{1}{\sqrt 2}(l^0 \pm l^3)$], and the soft partons
have momentum $(\lambda^2, \lambda^2, \lambda^2)Q$. The moment of
the structure function $F_1$ after the second stage matching has
the following form \cite{kim},
\begin{equation}
   F_1^N(Q) = C^2(Q^2/\mu_I^2) J_P(N, Q^2/\mu_I^2) \phi_N(\mu_I^2)
S_N(\mu_I^2) \ ,
\end{equation}
where various factors are defined as follows.

The jet function $J_P(N, \mu^2_I)$ is related to the absorbtive
part of the hard collinear quark propagator ${\cal G}_P$,
\begin{equation}
\langle 0|T\left[ W_n^\dagger\xi_n(z)
  \bar \xi_n W_n(0)\right]|0\rangle
   = i \frac{\not\! n}{\sqrt 2}\int \frac{d^4k}{(2\pi)^4}
   e^{-ikz} {\cal G}_P(k) \ .
\end{equation}
where $\xi_n$ is a collinear quark field and $W_n$ is a Wilson
line along the light-cone direction $\bar n$ $(n^2=\bar n^2=0,
\bar n \cdot n=1)$. A hard collinear quark has momentum $p + k$,
where $p$ is the so-called label momentum with $p^+\sim Q$ and $k$
is a hard residual momentum with components of order $Q(1,
\lambda^2, \lambda)$. Therefore, the virtuality of the
hard-collinear quark is $2p\cdot k \sim \lambda^2 Q^2$, consistent
with that of the hadron final state. The jet function has no
infrared divergences because the hadron final states are summed
over. However, it does have light-cone divergences which are
handled by the standard minimal subtraction method. An important
feature of the jet function is that it is only sensitive to
physics at scale $\mu_I^2$. In fact, at one-loop order, the jet
function reproduces the ${\cal M}_N$ function in the previous
section.

The soft contribution in Eq. (12) is defined in terms of the soft
Wilson lines \cite{chay1}: $Y_n(x)  = {\mathcal P}
\exp[-ig\int^x_{-\infty} ds n\cdot A_{\rm us}]$ and $\widetilde
Y_n(x)  = \overline{{\mathcal P}} \exp[-ig\int^\infty_{x} ds
n\cdot A_{\rm us}]$, where $A_{\rm us}$ are the so-called
ultra-soft gluons with momentum $Q(\lambda^2, \lambda^2,
\lambda^2)$ and ${\mathcal P}$ stands for path ordering [the sign
convention for the gauge coupling is
$D^{\mu}=\partial^{\mu}+igA^{\mu}$],
\begin{equation}
     S(x, \mu_I^2) = \frac{1}{N_c} \left\langle 0\left|{\rm Tr} \left[ {Y_{\bar
n}}^\dagger \tilde Y_n\delta\left(1-x+
     \frac{n\cdot i\partial}{n\cdot p}\right) \tilde Y^\dagger_n Y_{\bar n
     }\right]\right|0\right\rangle \ ,
\end{equation}
where the ratio in the delta function fixes the momentum of the
emitted gluon to be soft (of order $\lambda^2$). As such, the soft
factor is a non-perturbative contribution.

The collinear contribution,
\begin{equation}
    \phi(x,\mu_I^2) = \left\langle P\left|\bar \xi_{\bar n} W_{\bar n} \delta
    \left(x - \frac{n\cdot {\cal P}_+}{n\cdot p}\right)
    \frac{\not\! n}{\sqrt 2} W^\dagger_{\bar n} \xi_{\bar n}\right|P\right\rangle
\end{equation}
comes from collinear quarks and gluons with momentum
$(1,\lambda^2, \lambda)Q$ and $n\cdot {\cal P}_+$ is the total light-cone
momentum carried by the partons. In Ref. \cite{kim}, the collinear
contribution was identified as the usual Feynman parton
distribution. This, however, is incorrect because the soft gluons
with longitudinal momentum $(1-x)Q\sim \lambda^2$ in the proton
cannot be included in the collinear contribution according to the
definition of the collinear gluons in SCET. On the other hand, the
Feynman parton distribution contains a factorizable soft
contribution in the limit $x\rightarrow 1$ \cite{korchemsky, ji1}.
Further discussion on this issue will be made in the next
subsection as well as in the next section. The correct approach is
to combine the soft and collinear contributions together to get
the correct Feynman parton distribution in the $x\rightarrow 1$
limit.

Thus EFT arguments lead finally to the following refactorization, valid
when $(1-x)Q\sim \Lambda_{\rm QCD}$ at leading order in $1-x$,
\begin{equation}
    F_{1}^N(Q^2) = C^2(Q^2/\mu^2_I) J_P(Q^2/N\mu_I^2) q_N(\mu_I^2) + {\cal O}(1-x),
\end{equation}
where $q_N(\mu_I^2)$ is the moment of the quark distribution, and
the jet function is exactly the ${\cal M}_N$ function introduced in
the previous section. Although formally this factorization is made
at $\mu_I^2$, the product of factors is independent of it. The claim
of non-factorizability of DIS in this very regime in Ref. \cite{pecjak} was
criticized in \cite{manohar1}. On the other hand, the above result seems
consistent with that of Ref. \cite{neu} if used in the same regime.

The above factorization allows us to resum over large logarithms.
Since the physical structure function is $\mu$-independent, we can
take $\mu_I$ in the above expression to whatever value we choose.
For example, if one sets $\mu^2_I=Q^2$, all large logarithms are
now included in the jet function. One can derive a renormalization
group equation for $J_P$ \cite{manohar}. Solving this equation
leads to a resummation of large logarithms.

Alternatively, with the original scale $\mu_I^2$,  there are large
logarithms in $C$, which can be resummed using the renormalization
group equation and the anomalous dimension $\gamma_1$. The
resulting exponential evolution can be regarded as the evolution
of the jet function from scale $Q^2$ to $\mu_I^2$. The parton
distribution $q_N(\mu_I)$ runs from $\mu_I^2$ to a certain
factorization scale $\mu_F^2$ using the DGLAP evolution
\cite{dglap}. This running generates the logarithms from
initial-state parton radiations.

In the above refactorization, no scale $(1-x)Q$ appears explicitly
although soft and collinear gluons in SCET do have
reference to that scale. This explains that the factorization holds in
the region of our interest, namely, when $(1-x)Q\sim \Lambda_{\rm QCD}$.

\subsection{Collinear Contribution in SCET and Double Counting}

SCET is an operator approach designed to take into account
contributions from different regions in Feynman integrals.
Calculations in SCET are sometimes formal if without a careful
definition of regulators for individual contributions.
Occasionally, the regulators in different parts must be defined
consistently to obtain the correct answer. Otherwise, one can
easily lead to double counting. The same issue has been discussed
recently in Ref. \cite{zb}.

To see the need of consistent regulators in SCET, let us consider
the usual quark distribution in the proton. In the full QCD, the
quark distribution is defined as
\begin{equation}
  q(x) = \frac{1}{2}\int\frac{d\lambda}{2\pi}
    e^{i\lambda x}
    \langle P|\overline{\psi}(\lambda n)
      \not\! n e^{-ig\int^\lambda_0 d\lambda' n\cdot A(\lambda' n)}
  \psi(0)|P\rangle \ ,
\end{equation}
where $\psi(x)$ and $A(\lambda n)$ are full QCD quark and gluon
fields (here we use the vector $n$ with mass-dimension 1).  Now
suppose the nucleon is moving with a high momentum $Q$ in the $z$
direction. The quarks and gluons in the proton, in general, have
large $k^+$, and small momentum $k^-$ and $k_\perp$, in the sense
that they are collinear to the proton momentum. Therefore, one may
match the full QCD fields in the above expression to the
corresponding collinear fields in SCET.

However, the procedure is incomplete for wee gluons with
$(1-x)\sim \Lambda/Q$. Such a gluon has a soft longitudinal
momentum and is definitely included in the above gauge link. The
QCD factorization theorem shows that soft gluons do not make a
singular contribution to the parton density, but it does not
exclude the non-singular wee gluon contributions of type $
[\ln^k(1-x)/(1-x)]_+$. In fact, the wee gluon effect in the
$x\rightarrow 1$ limit can be factorized out into a soft factor
$S(x)$, which is responsible for the large-$x$ behavior of the
parton distribution \cite{korchemsky, ji1}. Therefore, in SCET it
is natural to express $q(x)$ in terms of the product of the soft
factor and true collinear gluon contribution.

In Ref.~\cite{kim}, the evolution equation was derived for the
collinear contribution $\phi(x)$ and is found to be the same
as the DGLAP evolution, even in the $x\rightarrow 1$ limit. This
could be the main motivation to identify $\phi(x)$ as $q(x)$.
However, the collinear gluons in the one-loop Feynman diagrams can
no longer be considered as ``collinear" if $(1-x)Q\sim \Lambda_{\rm QCD}$, and
must be subtracted explicitly. This subtraction was not made
through certain regulators and hence there is a double counting.
In fact, once the soft-gluons are subtracted, a collinear parton
jet shall not have singularity in the limit $x\rightarrow 1$.
Likewise, the calculation of the soft-factor in Ref.~\cite{kim}
should have a soft transverse-momentum cutoff to include just the
true soft gluons. Thus, the regulators in the soft and collinear
contributions must be made consistently to avoid double counting.
A consistent scheme of defining the soft and collinear
contributions for a parton distribution defined with off light-cone
gauge link can be found in Ref. \cite{korchemsky}. We will present another
example of a consistent regularization in the following section.

\section{ReFactorization: Intuitive Approach}

The EFT approach for refactorization is gauge-invariant and all
factors are defined at separate scales. The resummation is of the
simple renormalization-group type. However, the physical origin of
the large logarithms is not entirely transparent. For example, it
is well known in QED that the double infrared logarithms are
generated from soft radiations from jet-like lines. This is not
obvious in the EFT approach.

In the approach introduced by Sterman and others \cite{sterman},
the structure functions are factorized into different factors
which have clear physical significance, although each factor now
contains multi-scales. Explicit equations can be derived to bridge
the scales within these factors, which allow one to resum large
logarithms. The main shortcoming of this intuitive approach is the
introduction of a spurious scale $(1-x)Q$ in each factor, which
make them nonperturbative in the kinematic region of our interest.

In this section, we first briefly review Sterman's approach in
subsection A. We then introduce in subsection B a more general
factorization approach along this direction, which involves a
rapidity cutoff. With an appropriate cutoff, we
arrive at a picture similar to that of EFT. The example also shows
that a consistent soft subtraction must be made to obtain a
correct factorization.

\subsection{Sterman's Method}

Consider the lepton-nucleon DIS process in the Breit frame in
which the initial and final partons have similar momentum but move
in opposite directions. In the region $x\rightarrow 1$, the final
hadron state consists of a high-energy jet plus soft gluon
radiations. A so-called reduced diagram is shown in Fig. 1,
showing the space-time picture of the process. There are in
principle four different scales which are relevant: virtual photon
mass $Q^2$, final-hadron invariant mass $(1-x)Q^2$, the soft
parton energy radiated off the proton $(1-x)Q$, and finally the
genuine nonperturbative QCD scale $\Lambda_{\rm QCD}$.

\begin{figure}[t]
\label{fullcurrent}
\begin{center}
\includegraphics[width=3.0in]{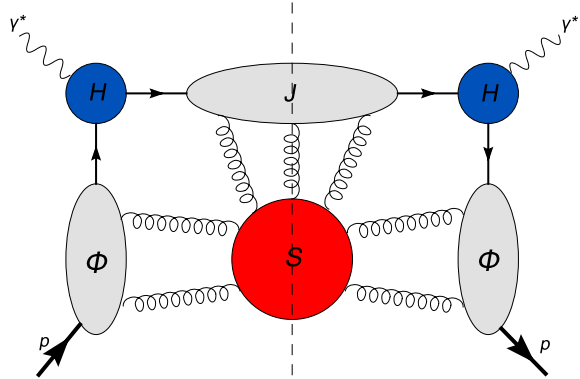}
\end{center}
\vskip -0.7cm \caption{The leading reduced diagram contributing to
the deep-inelastic structure function in $x\rightarrow 1$ regime.}
\end{figure}

According to the analysis in Ref. \cite{sterman}, the reduced
diagram can be factorized into various physically intuitive
contributions, and the structure function can be expressed as a
product of a soft factor, a final-state jet function, and a parton
distribution,
\begin{equation}
  F_1(x, Q^2) = H(Q^2)\int^1_x
  \frac{dy}{y}\phi(y)\int^{y-x}_0\frac{dw}{1-w}S(w)J(x,y,w)\ .
\end{equation}
The parton distribution $\phi(x)$ is not the usual gauge-invariant
one on the light-cone. Rather it is defined as
\begin{equation}
  \phi(y) = \frac{1}{2p^+}\int \frac{d\lambda}{2\pi} e^{-iy\lambda}
    \langle P|\overline{q}(\lambda n)\gamma^+ q(0)|P\rangle \ ,
\end{equation}
in $A^z=0$ gauge, or equivalently there are gauge links along the
$z$ direction going from the quark positions to infinity. Because
it is not truly gauge-invariant, it is frame-dependent. In
particular, it can depend on the soft parton energy $(1-x)Q$. This
parton distribution contains contributions of both collinear and
soft gluon radiations form the initial state quark, thus involving
double logarithms.

Likewise, the jet function is defined as
\begin{equation}
  J \sim  {\rm Disc} \int \frac{d^4x}{(2\pi)^4} e^{-ixl}
    \langle 0| T \left[ \overline{q}(x) q(0)\right]|0\rangle
\end{equation}
and normalized to $\delta(1-x)$ at the leading order. Once again
the jet is defined in the axial gauge and is frame dependent. In
particular, it depends on the infrared scale $(1-x)Q$ as well.
However, this jet function contains no true infrared divergences.
It accounts for the collinear and soft radiations from the jet
final state.

Finally, the soft function $S$ is defined as the matrix element of
Wilson lines first going along the $\bar n$ direction from
$-\infty$ along the light-cone, then going along $n$ direction to
$+\infty$. The collinear divergences are regularized by $A^z=0$
gauge. Therefore, it contains no true infrared divergences.

In the moment space, the refactorization appears in a simple form,
\begin{equation}
    F_1^N(Q^2) = H(Q^2/\mu^2) \phi_N(Q^2/\mu^2) S_N(Q^2/\mu^2) J_N(Q^2/\mu^2)\ ,
\end{equation}
and $H$ stands for the hard contribution which comes only from
virtual diagrams. Again, we emphasize that the factorization
follows intuitively from the space-time picture of the reduced
diagrams. However, one pays a price for this: the breaking of
Lorentz invariance and introduction of a new scale $(1-x)Q$. When
this scale becomes of order $\Lambda_{\rm QCD}$, as this is the
main interest of the paper, all factors becomes nonperturbative in
principle. On the other hand, the scale $(1-x)Q$ is spurious, it
should be cancelled out. It is unclear how this is achieved at the
nonperturbative level.

The physics of the above factorization is best seen through a
one-loop calculation. The parton distribution is
\begin{equation}
\phi(x) = \frac{\alpha_s}{\pi}C_F\left\{-\frac{1}{\epsilon}
P_{qq}(x)
  + \left[ \left(2D_1+ D_0\ln\frac{Q^2}{\mu^2}\right)
  -D_0- \left(\frac{\pi^2}{6} + 1\right) \delta(1-x)
   \right]\right\} \ ,
\end{equation}
where $P_{qq}(x)$ is the quark splitting function and
\begin{eqnarray}
D_i \equiv \left
[\frac{\ln^{i}(1-x)}{1-x}\right]_+\,\,\,\,\,i=0,1,2....
\end{eqnarray}
 Apart from the divergent term which is the same as the Feynman
parton distribution, there are extra constant terms which absorb
the soft-collinear gluon contribution. Some of these come from the
scale $(1-x)^2Q^2$. In moment space, we have
\begin{equation}
  \phi_N(x) = 1 + \frac{\alpha_s}{4\pi}C_F
  \left[-\frac{1}{\epsilon}(3-4\ln \overline{N})+\ln^2\frac{Q^2}{{\overline N}^2\mu^2}
       - 2 \ln \frac{Q^2}{{\overline N}^2\mu^2}- \ln^2\frac{Q^2}{\mu^2}
        +  2 \ln\frac{Q^2}{\mu^2} - 4\right] \ .
\end{equation}
One may view the double logarithmic terms as from the initial state
radiation. The large logarithms resulting from large scale
differences can be summed through an $x$-space evolution equation
\cite{sterman}.

The one-loop jet function is explicit finite,
\begin{eqnarray}
   J(x) &=& \delta(1-x)  + \frac{\alpha_s}{4\pi}C_F \left[4\left(D_1 +
D_0\ln\frac{Q^2}{\mu^2}\right) -
   4\left(2D_1 + D_0\ln\frac{Q^2}{\mu^2}\right)- 7D_0
   \right. \\ \nonumber && \left. +  \left(3
     -3\ln\frac{Q^2}{\mu^2}\right) \delta(1-x)\right] \ ,
\end{eqnarray}
which involves physics at both scales $(1-x)^2Q^2$ and $(1-x)Q^2$.
Both type terms generate double logarithms, corresponding to the
radiations from the jet. In moment space, the jet function
becomes,
\begin{eqnarray}
  J_N &=& 1 + \frac{\alpha_s}{4\pi}C_F
  \left[-\ln^2\frac{Q^2}{{\overline N}^2\mu^2}
       +2\ln \frac{Q^2}{{\overline N}^2\mu^2}
       +2\ln^2\frac{Q^2}{{\overline N}\mu^2}
       - 3\ln \frac{Q^2}{{\overline N}\mu^2} \right. \nonumber \\ && \left. -
       4\ln\frac{Q^2}{{\overline N}^2\mu^2}
        - \ln^2\frac{Q^2}{\mu^2} + 3
          - \frac{\pi^2}{3}
        \right]\ .
\end{eqnarray}
Again the large logs can be resummed through an evolution equation
for $J$ \cite{sterman}.

The soft function is also finite and at one-loop;
\begin{equation}
S(x) =  \delta(1-x)+ \frac{\alpha_s}{\pi}C_F \left[2D_0+
\left(\ln\frac{Q^2}{\mu^2}\right)\delta(1-x)\right]\  ,
\end{equation}
which contains physics at scale $(1-x)^2Q^2$, generating a single
logarithm. In the moment space, it is
\begin{equation}
    S_N = 1 + \frac{\alpha_s}{\pi} C_F \ln \frac{Q^2}{{\overline N}^2\mu^2} \
    .
    \label{soft}
\end{equation}
Here the collinear singularity is regulated by gauge fixing.

When summing over the jet, parton distribution and soft
contribution, the soft scale $(1-x)^2Q^2$ dependence cancels. We
are left with only the physical scale $(1-x)Q^2$. All the factors
introduced above are sufficient to factor away the singular
contributions in the structure function at $x\rightarrow 1$ limit.
In fact, at one-loop
\begin{equation}
  F_1^{(1)}(x,Q^2)
   - (\phi^{(1)}(x) + J^{(1)}(x) + S^{(1)}(x)) = \frac{\alpha_s}{2\pi} C_F
   \left[-4+\ln\frac{Q^2}{\mu^2}\right]\delta(1-x)\ ,
\end{equation}
which contains only the $\delta$-function singularity. Therefore, the
large double logarithms have been absorbed either into the parton
distribution or the jet function. This is, in fact, the purpose of
the intuitive refactorization approach: The double logarithms
from the initial and final state radiations are made
explicit through factorization.

However, because of the presence of the extra scale $(1-x)Q$, the
above refactorization is not very useful in the region of our
interest because all factors, except $H$, become nonperturbative.

\subsection{Alternative Regulator, Consistent Subtraction and Relation to
SCET Factorization}

In defining various contributions in Sterman's approach, the
gauge choice $A^z=0$ is made, or equivalently gauge links along
the $z$-direction are added to operators to make them gauge
invariant. This choice of a non-light-like gauge can serve in addition
as a regulator for collinear divergences arising from
gauge links going along the light-cone direction, as can be seen
from the one-loop the soft factor, Eq.~(\ref {soft}).

In this subsection we present an alternative method to arrive at the
correct factorization formula with factors that are manifestly
gauge invariant.
We regulate collinear divergences by choosing gauge links slightly
off the light-cone (for more discussion of off-light-cone gauge
links see \cite{korchemsky, coll,ji1}.) The direction of a gauge link
supports a finite rapidity which can serve as a rapidity cutoff,
thereby avoiding light-cone singularities which appear in
calculations of scaleless quantities like the parton distribution
and which cannot be regularized by dimensional regularization
\cite{coll2}. We show that the factorization theorem proposed
here is obtained only after proper subtractions of soft factors
are made. By choosing the rapidity
parameter $\rho$ appropriately, we can eliminate the intermediate
scale $(1-x)Q$ and arrive at a factorization similar to that of
EFT.

Let ${\tilde v}=({\tilde v}^+,{\tilde v}^-,0)$ with ${\tilde
v}^+\gg {\tilde v}^-$ and $v=(v^+,v^-,0)$ with $v^-\gg v^+$ with
$\rho \equiv {\tilde v}^+/{\tilde v}^-=v^-/v^+$. It is assumed
below that the incoming quark is collinear in the $z$ direction
with momentum $p_1=(Q/{\sqrt 2},0,0)$ and the outgoing quark is
collinear in the $-z$ direction with $p_2=(0,Q/{\sqrt 2},0)$, and
we denote $p^+=p_1^+=p_2^-=Q/{\sqrt 2}$. We define the soft factor
as
\begin{eqnarray}
  S(1-x)&=&\int \frac{d\lambda}{2\pi} e^{i\lambda (1-x)p^+}
    \langle 0|{\rm Tr}[Y^\dagger_{\tilde v}(0,-\infty;\lambda {\tilde
v})Y^\dagger_v(\infty,0;\lambda v)
\nonumber \\
    && \times Y_v(\infty, 0; »0)Y_{\tilde v}(0,-\infty ; 0)]|0\rangle
    \frac{1}{N_c}
\end{eqnarray}
where $Y_v$ is
\begin{equation}
   Y_v(a,b; \xi) =  \mathcal{P}\exp\left(-ig\int^a_b d\lambda' v\cdot
A(\lambda'
   v+ \xi)\right) \ ,
\end{equation}
 Similar definition holds for
$Y_{\tilde v}$. Thus the soft factor depends on two off-light-cone
Wilson lines
  in the directions of  $v$ and ${\tilde v}$.
This definition of the soft factor has no collinear divergences.
On the other hand, if we take one of the Wilson lines on the
light-cone, the resulting light-cone divergence may be considered
as the collinear divergence. Then the $S$ factor will include a
collinear contribution as discussed in \cite{coll}. However, here
we are interested in the soft factor that is not contaminated with
collinear divergences.

The final state jet function is defined as follows:
\begin{equation}
\tilde J(x, Q) = \int \frac{d\lambda}{2\pi} e^{i\lambda (1-x)p^+}
\langle 0 | W^\dagger_{v}(\infty,0;\lambda {\tilde v})
\psi(\lambda {\tilde v}) \overline \psi(0) W_{
v}(\infty,0;0)|0\rangle\ , \label{jet}
\end{equation}
where it involves a Wilson line in the $\tilde v$ direction which
is taken to be in the (almost) conjugate direction to the out-going partons.
It is given by:
\begin{eqnarray}
W_v(a,b;\xi)={\mathcal P}\exp\left(-ig\int^a_b d\lambda' {\tilde
v}\cdot A(\lambda'
   {\tilde v}+ \xi)\right).
   \end{eqnarray}
Finally we define the parton distribution function as
\begin{eqnarray}
\tilde \phi (x)  &=& \frac{1}{2p^+}\int \frac{d\lambda}{2\pi}
e^{i\lambda x\!p^+} \langle P | \overline{\psi}(\lambda v) W_{
{\tilde v}}^\dagger(\infty,0;\lambda v) \nonumber \\ && \times
\gamma^+ W_{\tilde v}(\infty,0;0)\psi (0)|P\rangle \ , \label{pdf}
\end{eqnarray}
where the Wilson line is taken along the (almost) conjugate
direction of the incoming partons, i.e., in the  $v$ direction.
Both the jet and parton distribution are in principle defined to absorb just
the collinear gluon contributions. Although this can be done by
cutoffs in loop integrals, it is difficult to achieve in an
operator approach. In fact, it will be clear later that both
$\tilde J(x)$ and $\tilde \phi(x)$ do contain soft contributions as well,
which must be subtracted explicitly.

Let us consider the factorization of the DIS structure function at
one-loop using the above definitions of the factors. We first
calculate the soft factor, jet function, and parton distribution
in pQCD. The one-loop soft factor is
\begin{eqnarray}
S(1-x) & = & \delta(1-x)+\frac{\alpha_s}{2\pi}C_F
\left(-2+\frac{\rho^2+1}{\rho^2-1}\ln\rho^2\right)
\left(2D_0+\ln\frac{Q^2}{\mu^2}\delta(1-x)\right)\nonumber \\
& \approx & \delta(1-x)+\frac{\alpha_s}{2\pi}C_F
\left(-2+\ln\rho^2\right)
\left(2D_0+\ln\frac{Q^2}{\mu^2}\delta(1-x)\right) \ ,
\end{eqnarray}
where the second line is obtained by taking the large $\rho^2$ limit.
The result does not have any soft and collinear divergences. It does have
an ultraviolet divergence coming from the cusp of the Wilson lines, which
has been subtracted minimally in dimensional regularization. In moment space,
\begin{eqnarray}
S_{N}=1+\frac{\alpha_s}{2\pi}C_F \left[2-\ln \rho^2 \right]\ln
\frac{\mu^2{\overline N}^2}{Q^2}  \ .
\end{eqnarray}
The UV-subtracted soft factor obeys the following renormalization group
equation
(RGE),
\begin{equation}
    \mu \frac{\partial S(1-x,\mu^2)}{\partial \mu} = 2\gamma_S ~S(1-x,\mu^2)
\ ,
\end{equation}
where the anomalous dimension is
\begin{equation} \gamma_{S} = \frac{\alpha_s}{2\pi}C_F
(2-\ln \rho^2)\ ,
\end{equation}
which depends on the rapidity cutoff $\rho$ and is related to the
so-called cusp anomalous dimension \cite{kor}.

The jet function has no infrared divergences either. At one-loop,
\begin{eqnarray}
\tilde J(x,Q)&=& \delta(1-x)+\frac{\alpha_s}{4\pi}
C_F\left[\left(3-2\pi^2-3\ln \frac
{Q^2}{\mu^2}+4\ln\rho-2\ln^2\rho\right ) \delta(1-x)\right.
\\ \nonumber
&&-7D_0-4D_1+4\ln\rho\, D_0\Big] \ .
\end{eqnarray}
In moment space, it has a particularly simple form,
\begin{eqnarray}
\tilde J_N=1+ \frac{\alpha_s}{4\pi}C_F\left\{3\ln \frac{{\overline
N}\mu^2}{Q^2} -2[\ln({\overline
N}\rho)-1]^2+5-\frac{7}{3}\pi^2\right\} \ .
\end{eqnarray}
The wave function renormalization brings in the scale-dependence
of the jet function, which therefore obeys the following RGE,
\begin{equation}
    \mu \frac {\partial \tilde J_N}{\partial \mu} = \gamma_{J,\mu} \tilde
J_N \ ,
\end{equation}
with
\begin{equation}
\gamma_{J,\mu}=\frac{3\alpha_s}{2\pi}C_F \ ,
\end{equation}
which is the anomalous dimension of the quark field in axial gauge
\cite{flo}.

The parton distribution at one-loop is,
\begin{eqnarray}
\tilde \phi(x) & = & \delta(1-x)+ \frac{\alpha_s}{4\pi}
C_F\left\{-\frac{1}{\epsilon}P_{qq}(x)
-\left(4+\frac{2\pi^2}{3}\right)\delta(1-x)
  \right.\nonumber\\
&
&\left.+\left(4\ln\frac{Q^2}{\mu^2}+4\ln\rho-4\right)D_0+8D_1\right\}
\ ,
\end{eqnarray}
where the $1/\epsilon$ pole comes from collinear divergences.
In moment space we have
\begin{equation}
\tilde \phi_N=\frac{\alpha_s}{4\pi}\left\{
-\frac{1}{\epsilon}\Big[3-4\ln {\overline N}\Big] +4\left[-1+\ln
{\overline N} +\ln^2 {\overline N} -\ln\overline{N}\ln\rho-\ln
{\overline N}\ln\frac{Q^2}{\mu^2}\right]\right\} \ .
\end{equation}
Its UV divergences come from wave function corrections and have been
subtracted minimally, and therefore its evolution in $\mu^2$ is
the same as that for the jet function.

Because $\tilde \phi$ and $\tilde J$ contain the soft contribution
as well, the structure function $F_1$ cannot be factorized into
$\tilde \phi\otimes \tilde J \otimes S$ multiplied by a hard
contribution, where $\otimes$ is a convolution operator in
$x$-space. Instead, one must define the soft-subtracted version of
the parton distribution $\phi = \tilde \phi/S$ and jet $J =\tilde
J/S$. Then the factorization reads:
\begin{eqnarray}
F_1(x, Q^2) &=& H(Q^2) \otimes J(x,Q^2/\mu^2) \otimes \phi(x, Q^2/\mu^2)
   \otimes S(x, Q^2/\mu^2) + {\cal O}(1-x) \nonumber \\
   &=& H(Q^2)\otimes \tilde J(x, Q^2/\mu^2)\otimes \tilde \phi(x, Q^2/\mu^2)/S(x, Q^2/\mu^2)
   + {\cal O}(1-x)   \ ,
\end{eqnarray}
where $H(Q^2)$ is hard contribution independent of $x$.
This can be easily checked at one-loop level by calculating $H$,
\begin{eqnarray}
H^{(1)}(Q^2) & = & F_{1}^{(1)}(x, Q^2)-\tilde J^{(1)}(x, Q^2)
- \tilde \phi^{(1)}(x, Q^2) + S^{(1)}(x, Q^2)\nonumber\\
& = & \left[1+\frac{\alpha_s}{4\pi}\left(2\ln\frac{Q^2}{\mu^2}+4
\ln\frac{Q^2}{\mu^2}\ln\rho+2\ln^2\rho-4\ln\rho-8+2\pi^2\right)\right]\delta(1-x)
\end{eqnarray}
which is indeed independent of $D_i(x)$. All singular contributions
of type $1/(1-x)_+$ have been subtracted from the structure
function $F_1$ by the jet, soft factor and parton distribution.
We emphasize that this is possible only when
the soft contributions to the jet and parton distributions have been
subtracted first. The anomalous dimension of the hard part is
\begin{eqnarray}
\gamma_{H}& \equiv &
\frac{\mu}{H}\frac{dH}{d\mu}=-\frac{\alpha_s}{\pi}C_F\big(2\ln\rho+1\big)
\ ,
\end{eqnarray}
which is $\rho$-dependent.

The above factorization uses the rapidity cutoff parameter $\rho$,
which has the similar role as the renormalization scale
$\mu^2$: Every factor is a function of it, but the product has no
dependence. Therefore, one can get different pictures of
refactorization by choosing different value of $\rho$. For
instance, if one takes $\rho\rightarrow \infty$, all gauge links
move back to the light-cone. Here collinear divergences shows up
in different factors which have to be subtracted beforehand to
yield a meaningful factorization. Distribution $\tilde \phi$
corresponds to the physical quark distribution $q(x)$.
The subtraction of the soft contribution in $\tilde J(x)$ ensures $J(x)$
have collinear contributions only. Thus factorization can be written as
\begin{equation}
F_1(x, Q^2) = H(Q^2) \otimes J(x,Q^2/\mu^2) \otimes \tilde \phi(x, Q^2/\mu^2)
 + {\cal O}(1-x)
\end{equation}
which is heuristically similar to Eq. (15)

One can also take $\rho =-1$. Then $\tilde J(x) = J_{\rm st}(x)$, and
$\tilde \phi(x) = \phi_{\rm st}(x)$ and $S(x) = S_{\rm st}^{-1}(x)$, where quantities
with subscripts ``st" refer to those in the previous subsection. Eq. (44)
then reproduces the factorization Eq. (20) from Ref. \cite{sterman}.

We can also make contact with the EFT approach
by taking $\rho$ to be small although it shall be considered as large in principle.
Consider the moment of $\phi(x)= \tilde \phi(x)/S$,
\begin{eqnarray}
\phi_N=1+ \frac{\alpha_s}{4\pi}
C_F\left\{-\frac{1}{\epsilon}\left[3-4\ln\overline{N}\right]+4\left[1-\ln({\overline
N}\rho)\right] \ln\frac{Q^2}{{\overline N}\mu^2}-4\right\} \ ,
\end{eqnarray}
where the finite part depends on a single logarithm
$\ln\frac{Q^2}{\bar{N}\mu^2}$, i.e., the scale $(1-x)Q$. In
the leading-logarithmic approximation, if $\rho$ is taken to be
$1/\overline{N}$, the distribution $\phi(x)$ no longer depends on
scale $(1-x)Q$. The anomalous dimension of the distribution
$\phi(x)$ is
\begin{equation}
\gamma_{\phi, \rho}\equiv\frac{\mu}{\phi_N}
\frac{d\phi_N}{d\mu}=\frac{\alpha_s}{2\pi}C_F(-1+4\ln\rho) \ .
\end{equation}
With $\rho\sim 1/\overline{N}$, the evolution equation for $\phi(x)$ is
similar to that of the light-cone quark distribution $q(x)$.

Now let us examine the refactorization in the following form,
\begin{equation}
    F_1^N(Q^2) = H(Q^2/\mu^2,\rho) \otimes \tilde J_N(Q^2/\mu^2,\rho )
    \otimes \phi_N(Q^2/\mu^2, \rho)|_{\rho~\sim~1/\overline{N}} + {\cal O}(1-x) \ ,
\end{equation}
Taking $\mu^2 = \mu_I^2 = Q^2/N$, the jet factor $\tilde J_N$
in Eq. (39) does not contain any large logarithms. The hard factor $H(Q^2/\mu^2)$
contains large logarithms that can be resummed.
The resummation generates exactly the evolution of the matching
coefficient $C^2$ for the product of effective currents in SCET.
Therefore, the above form of factorization exactly reproduces the
SCET result in Eq. (15).

\section{Summary}

In this paper, we considered deep-inelastic scattering in a region
$(1-x)Q\sim \Lambda_{\rm QCD}$ where the standard pQCD
factorization is not supposed to work. We argued, however, there
is nothing that invalidates it in the new regime in leading order
$1-x$ because the Lorentz invariant factorization does not involve
this soft scale in the sense that there are no new infrared
divergences associating with this scale.

We then discussed refactorization of the coefficient function. The
EFT approach maintains Lorentz invariance and hence allows a
form of refactorization which is valid in the new regime. However,
in the traditional approach in which jets and parton distributions
are defined to take into account explicitly the double-logarithmic
soft radiations, the scale $(1-x)Q$ does appear in various
factors, making them nonperturbative in nature. We consider a more
general factorization in this spirit which involves a rapidity
cutoff. We showed how the EFT result can be reproduced through
choices of this cutoff. The example also shows how to make
consistent subtraction of the soft contribution in collinear matrix elements.

\section*{Acknowledgments}
We acknowledge support of the U. S. Department of Energy via grant
DE-FG02-93ER-40762. X. J. is partially supported by the China
National Natural Science Foundation and Ministry of Education.

\end{document}